\documentclass[runningheads]{llncs}
\usepackage{graphicx}
\usepackage{amsmath,amssymb,amsfonts}
\usepackage{lipsum} 
\usepackage{algorithmic}
\usepackage{graphicx}
\usepackage{textcomp}
\usepackage{multicol} 
\usepackage{multirow} 
\usepackage[export]{adjustbox} 
\usepackage{array,threeparttable,siunitx} 
\usepackage{tabularx} 
\usepackage{booktabs} 
\usepackage{color}
\usepackage{xspace}
\usepackage{url}
\usepackage[hang,flushmargin]{footmisc} 
\usepackage[colorlinks=true]{hyperref} 

\usepackage{csquotes} 
\usepackage{subfigure}
\usepackage[misc,geometry]{ifsym} 
\begin{document}
\title{Attributes affecting user decision to adopt a Virtual Private Network (VPN) app\vspace{-2ex}}
\titlerunning{Attributes affecting user decision to adopt a VPN app (PREPRINT)}
%
\author{Nissy Sombatruang\inst{1,2} $^{(\textrm{\Letter})}$
Tan Omiya\inst{3}
Daisuke Miyamoto\inst{3}
M. Angela Sasse\inst{1,5}
Youki Kadobayashi\inst{3}
Michelle Baddeley\inst{6}}
\authorrunning{N. Sombatruang et al.}
%
\institute{University College London, UK
\and National Institute of Information and Communications Technology, Japan
\email{nissy@nict.go.jp} 
\and Nara Institute of Science and Technology, Japan\\
\email{\{tan.omiya1, youki-k\}@is.naist.jp}
\and University of Tokyo, Japan\\
\email{daisu-mi@nc.u-tokyo.ac.jp}
\and Ruhr-University Bochum, Germany\\
\email{martina.sasse@ruhr-uni-bochum.de}
\and University of Technology Sydney, Australia\\
\email{michelle.baddeley@uts.edu.au}
\vspace{-2ex}}

\maketitle       

\begin{abstract}
A Virtual Private Network (VPN) helps to mitigate security and privacy risks of data transmitting on unsecured network such as public Wi-Fi. However, despite awareness of public Wi-Fi risks becoming increasingly common, the use of VPN when using public Wi-Fi is low. To increase adoption, understanding factors driving user decision to adopt a VPN app is an important first step. This study is the first to achieve this objective using discrete choice experiments (DCEs) to elicit individual preferences of specific attributes of a VPN app. The experiments were run in the United Kingdom (UK) and Japan (JP). We first interviewed participants (15 UK, 17 JP) to identify common attributes of a VPN app which they considered important. The results were used to design and run a DCE in each country. Participants (149 UK, 94 JP) were shown a series of two hypothetical VPN apps, varying in features, and were asked to choose one which they preferred. Customer review rating, followed by price of a VPN app, significantly affected the decision to choose which VPN app to download and install. A change from a rating of 3 to 4-5 stars increased the probability of choosing an app by 33\% in the UK and 14\% in Japan. Unsurprisingly, price was a deterrent. Recommendations by friends, source of product reviews, and the presence of in-app ads also played a role but to a lesser extent. To actually use a VPN app, participants considered Internet speed, connection stability, battery level on mobile devices, and the presence of in-app ads as key drivers. Participants in the UK and in Japan prioritized these attributes differently, suggesting possible influences from cultural differences.

\end{abstract}

\noindent\fbox{\begin{minipage}{\dimexpr\textwidth-2\fboxsep-2\fboxrule\relax}
\scriptsize
First published in the 22\textsuperscript{nd} International Conference on Information and Communications Security (ICICS 2020), Copenhagen, Denmark.
\end{minipage}}

\section{Introduction} 
\label{sec:intro}
VPN provides an encrypted channel for data transmission. It mitigates the privacy and security risks of user data when using unsecured networks such as public Wi-Fi. Although these risks can be mitigated by encrypting data at the application or at the network layer, the reality today is not all apps, websites, and Wi-Fi access points perfectly encrypt all data they transmit. Until that happens, encouraging users to use a VPN app is sound approach.

More than 200 VPN apps are available on both Google Play and Apple App Stores each\footnote{As of May 2020.}. However, VPN adoption for personal use and for security and privacy purpose is relatively low. The 2017 Norton Wi-Fi Risk Report examining consumers' public Wi-Fi practices showed that only 25\% of 15,532 survey respondents from 15 countries mentioned they used VPN \cite{symantec} --- which was worrying, given that three-fourths of participants put their data at risk. More concerning is that 80\% of these participants also admitted having used public Wi-Fi for email and online banking \cite{symantec}. As the use of public Wi-Fi and cyber risks continues to grow \cite{huffington,csoonline,sombatruang2016people,sombatruang2018continued,fsecure,ali2019privacy,chen2010side}, understanding factors affecting the decisions to adopt a VPN app is key to identifying suitable strategies to promote its uses. 

Previous studies examining drivers for VPN adoption focused on the transparency of VPN service \cite{khan2018empirical}, the awareness and trust of VPN \cite{ghaoui2017policy}, and the security and privacy of the VPN apps \cite{ikram2016analysis,appelbaum2012vpwns,perta2015glance,zhang2017oh}. However, none examined the effects of a VPN app's attributes on individuals decision to adopt it. Our study aimed to bridge this gap in the knowledge. Specifically, we investigated the attributes affecting the decisions to a) download and install an app --- referred to as \textit{the uptake} hereafter --- and b) actually use a VPN app. 

We conducted a semi-structured interviewed with participants in the UK (15) and Japan (17) to identify common attributes of a VPN app which they considered important for the adoption. The results were then used to design and run DCEs, the quantitative method to elicit individual preferences to specific attributes of a product by exploring the full landscape of potential choices, with participants in the UK (149) and Japan (94). Our findings showed that several attributes of a VPN app significantly affected the decisions to adopt it. However, participants in the two countries prioritized these attributes differently and some of these drivers stem from the herding attitude and resource preservation heuristics. These insights would help VPN app providers to design a more desirable VPN app and government agencies keen to promote online safety to develop a more workable awareness campaign to promote the use of VPN.

In summary, our contributions are as followed. We investigated drivers for a VPN app adoption in the UK and Japan, being the first to use DCEs. We showed that several VPN app's attributes affecting the decision to adopt an app, and that preferences for these attributes were not always universal in nature and that some of them stemmed from biases in decision-making.

\section{Related work} 
\label{sec:related}
\subsection{Current state of VPN uptake and usage}
Various sources reported different statistics of the current state of VPN uptake and usage. Norton \cite{norton}, a cyber security company, reported in 2016 that 16\% of people in the UK used a VPN when using public Wi-Fi. In 2017, the Norton Wi-Fi Risk Report examining consumers' public Wi-Fi practices showed that 25\% of 15,532 survey respondents from 15 countries (UK and JP included) mentioned they used VPN \cite{symantec}. Although the trend was upward, the 25\% was relatively low, given that three-fourths of these participants put their data at risk and that 80\% of them admitted having used public Wi-Fi for email and online banking \cite{symantec}. 

Another survey in 2017, by YouGov \cite{yougov}, reported that 16\% of British adults used either a VPN or proxy server --- mostly for accessing contents not available to them locally (48\%) but also for extra security (44\%) and for extra privacy (37\%). From the standpoint of security and privacy of user data, the reported 16\% was fairly low. Published statistics for VPN usage in Japan were more difficult to find. Nonetheless, VPNmentor \cite{vpnmentor} reported that Japan ranked amongst the countries\footnote{Along with Australia, Poland, Canada, Netherlands, and France.} with the lowest use of VPN. 

\subsection{Factors affecting VPN uptake and usage}
Previous studies investigating drivers for VPN adoption for personal use are few. Using desk research and an interview with a technical expert from the Dutch National Cyber Security Centre, Ghaoui \cite{ghaoui2017policy} identified obstacles that had led to low adoption in the country including the lack of awareness of VPN, difficulties in comparing VPN apps, and distrust of VPN providers. Similarly in Japan, Kaspersky \cite{kaspersky} reported in 2019 a lack of awareness of VPN in the country. Of the 624 survey participants, 35\% said they knew about VPN. The issue of the difficulties in comparing VPN apps identified in the Netherlands \cite{ghaoui2017policy} is also likely to apply elsewhere. There are more than 200 VPN apps on Google Play and Apple App Store today\footnote{As of May 2020.}; comparing these numerous apps is challenging even for someone with a technical background.

One possible reason for low VPN adoption may lie in the potential security and privacy flaws of the apps themselves. A number of studies provided  evidence that many VPN apps were prone to several risks: de-anonymization attacks \cite{appelbaum2012vpwns}, traffic leakage \cite{perta2015glance,khan2018empirical}, insecure VPN tunnelling protocols and DNS traffic leakage \cite{ikram2016analysis}, VPN traffic de-encryption and Man-in-the-Middle attack \cite{zhang2017oh}, and lack of transparency of VPN services \cite{khan2018empirical}. However, ordinary users are unlikely to truly understand these technical issues; hence, arguably, these issues may not affect the decisions to adopt VPN. 

Previous studies shed some light on possible reasons for the low VPN adoption. However, none of them examined the attributes of a VPN app that could influence the adoption. Our study aimed to address this gap. 

\section{Methodology} \label{sec:methodology} 
\subsection{Background of Discrete Choice Experiment (DCE)} 
DCE is an attribute-based survey method for measuring utility (a level of satisfaction)~\cite{ryan2007using}. Many areas of study, from marketing to health care, use DCE --- but fewer so in cyber security. A DCE is used to elicit individual preferences of specific attributes of a product or service. Hence, the DCE can yield several useful insights such as guiding the design of a product or marketing strategies~\cite{sas}. 

In a DCE, participants are asked to state their preferences for a hypothetical, yet generally realistic product. This usually involves presenting them with a series of hypothetical choice sets. Each choice set comprises of two (or more) competing products (e.g., product A and B) having the same set of attributes (e.g., price and customer review rating). However, the value of at least one (or more) attributes vary (e.g., A is \pounds{0.99} and B is \pounds{1.99}). Participants are asked to choose the choice they prefer (e.g., A or B). The varying attributes' value allows us to observe how participants perceive the importance of each attribute, and to identify key attributes affecting decision-making, accordingly.

The DCE, based on Lancaster's economic theory of value \cite{lancaster1966new}, assumes that participants derive utility from the underlying attributes of the product utility --- generally referred to as the \textit{main effects} --- (eq. \ref{eq:1}) and participants select the choice which maximizes their utility (eq. \ref{eq:2}) \cite{ryan2007using,ryan2008discrete}. 

\begin{equation} \label{eq:1}
U_{in} = V(X_{in}, \beta) + \varepsilon_{in}  
\end{equation}

\noindent Where $U_{in}$ is the latent utility of choice $i$ as perceived by the individual $n$; $V(X_{in}, \beta)$ is an explainable component, specified as a function of the attributes of choice $i$; and a random (unexplainable) component $\varepsilon_{in}$ is the unmeasured variation in preferences which could be caused by factors such as unobserved attributes, or measurement errors \cite{ryan2007using}.

Participant \textit{n} will choose choice $i$ if it maximizes their utility among all $j$ alternatives included in choice set $C_{n}$. That is,

\begin{equation} \label{eq:2}
U_{in} > U_{jn} \forall j\neq i \in C_{n}
\end{equation}

\noindent Where $U_{in}$ is the latent utility of choice $i$ as perceived by individual $n$; and $U_{jn}$ is the latent utility of the alternative choice $j$ as perceived by individual $n$.  

Since $\varepsilon_{in}$ and $\varepsilon_{jn}$, are unobservable and unmeasurable, it is not possible to conclude exactly whether $\varepsilon_{in}>\varepsilon_{jn}$; hence, the choice outcome can only be determined in terms of probability (eq. \ref{eq:3}) \cite{ryan2007using,mcfadden1973conditional}. That is,

\begin{equation} \label{eq:3}
P_{in} = Pr(U_{in} > U_{jn} \forall j \neq i \in C_{n})
\end{equation}

\noindent Where $P_{in}$ is the probability of participant $n$ selecting choice $i$; and $Pr$ is the probability of $U_{in} > U_{jn}$.

The choice statistical model can also be written as,

\begin{equation} \label{eq:4}
P_{in} = \exp(U_{in})/\Sigma_{j \in C_{n}} \exp(U_{jn})
\end{equation}

Designing a DCE involves several steps \cite{johnson2013constructing}. The first step is usually identifying the key attributes and their values --- hereafter referred to as \textit{attribute levels}. There could be an infinite number of attributes and attribute levels. However, not all of them are key in driving decision-making. Several techniques can help to narrow them down such as using focus groups, or user interviews (used in this study) \cite{ryan2007using}. Once the key attributes and attribute levels have been identified, the next step is designing the choice sets and the user interface of the actual experiment. When the experiment has been tested and finalized, participants are recruited, data collection commences, and the choice analysis follows.

\subsection{Experimental design} 
\subsubsection {Attributes and attribute levels identification}
We conducted user interviews to identify a common set of attributes and attribute levels of a VPN app likely to or have influenced the decisions to adopt a VPN app.

\paragraph{Recruitment} In the UK, we advertised our study on noticeboards at public space and via online media. In Japan, we advertised our study via student and staff mailing lists and verbally in classrooms (we were only permitted to conduct the study with students and staff). Eligible participants were restricted to residents of the UK/Japan, age at least 18 years old, all of whom had a smartphone, and used public Wi-Fi at least from time to time. A total of 32 participants (15 UK and 17 JP) were recruited from mixed demographics (Appendix: Table ~\ref{table:table6}). Each participant in the UK was awarded a \pounds{10} gift voucher. In Japan, for two of the institutions, each participant was awarded a \textyen1000 gift voucher. Participants at another institution, however, were recruited on a voluntary basis.

\paragraph{Interview structure} We conducted a one-hour face-to-face semi-structured interview with each participant. The questions set the scene by asking participants about their use of public Wi-Fi and the risks they perceived, and their prior experience with VPN and/or a VPN app. If they had never heard of or used VPN before, we explained and demonstrated how it works. We then asked them about attributes of a VPN app which would or have influence(d) them to download and install and actually use it. Interview questions were the same for the interviews in the UK and Japan. However, interviews in Japan was conducted in Japanese. Interview sessions were audio-recorded, transcribed, and, for the interviews in Japan, translated to English for data analysis.

\paragraph{Analysis of user interviews} We analyzed the transcriptions to identify common attributes of a VPN app deemed by participants in each country as crucial. This involved two steps. First, each transcription was reviewed manually and the attributes of a VPN app which each participant said were important for the uptake and the actual uses of the app were recorded. Since participants did not always use the same terminology for the same attributes, we standardized the attribute names and grouped them manually (where possible).

Next, for each country, we used Microsoft Power BI's text analysis function to analyse the frequency of each attribute i.e. how many participants considered the attributes to be important. Attributes which were mentioned by many participants ($min_{uptake}$: $UK = 5$, $JP = 10$; $min_{actualuse}$: $UK = 12$, $JP = 9$) were chosen to be tested in the DCEs. However, we also included the presence of the in-app ads in the actual uses (of a VPN app) part of the experiment for Japan despite not meeting the minimum frequency. The rationale was this attribute was included in the uptake part of the experiment; hence, we wanted to test whether the effect of this attribute persisted in the actual use of a VPN app.

To identify attribute levels, insights drawn from the analysis of the interview transcriptions and desktop research were used. A summary of attributes and attribute levels being tested in the DCEs is in Table ~\ref{table:table1}.

\begin{table}[tp]
\caption{A summary of attributes and attribute levels tested}
\centering
\scriptsize
\begin{threeparttable}
\begin{tabular}{|c|c|c|}
\hline
\multicolumn{3}{|c|}{\textbf{Uptake of a VPN app}}\\ 
\hline
\textbf{Attribute} &\textbf{UK} &\textbf{JP}\\
\hline

\multirow{4}{15em}{Price} &Free &Free\\
&\pounds{0.99}/one-off &\textyen100/one-off\\
&\pounds{4.99}/month &\textyen500/one-off\\
& &\textyen1,000/month\\
\hline

\multirow{3}{15em}{App review rating} &Good (4-5 stars) &Same as the UK\\
&Moderate (3 stars) &\\
&Bad (1-2 stars) &\\
\hline

\multirow{3}{15em}{No. of app downloads} &$>$100,000 &$=>$1000 downloads\\
&10,000-100,000 &$<$1000 downloads\\
&$<$10,000 &\\
\hline

\multirow{2}{15em}{User interface} &Professional-looking &n/a\\
&Amateur-looking &\\
\hline

\multirow{2}{15em}{Recommended by friends} &Yes &n/a\\
&No &\\
\hline

\multirow{2}{15em}{Source of app review} &n/a &App store\\
& &Tech blog/Websites\\
\hline

\multirow{2}{15em}{Installation time} &n/a &$=>$5 mins\\
& &$<$5 mins\\
\hline

\multirow{2}{15em}{In-app ads} &n/a &Yes\\
& &No\\
\hline

\multicolumn{3}{|c|}{\textbf{Actual use of a VPN app}}\\
\hline
\textbf{Attribute} &\textbf{UK} &\textbf{JP}\\
\hline

\multirow{3}{15em}{No. of dropped connections/hr} &1-2 &Same as the UK\\
&3-4 & \\
&$>$4 &\\
\cline{1-2}

\multirow{3}{15em}{Internet speed when using VPN} &10-20\% slower &\\
&21-30\% slower & \\
&$>$30\% slower &\\
\cline{1-2}

\multirow{4}{15em}{Battery level on mobile phone} &75-100\% &\\
&50-74\% & \\
&25-49\% &\\
&$<$25\% &\\
\cline{1-3}

\multirow{3}{15em}{VPN initiation method} &Automatic &Automatic\\
&On-demand -- via app &On-demand \\
&On-demand -- via task bar &\\
\cline{1-3}

\multirow{2}{15em}{In-app ads} &n/a &Yes\\
& &No\\
\hline

\end{tabular}
\end{threeparttable}
\label{table:table1}
\end{table}

\subsubsection{Choice set design}
The main objective of this step is to decide how many combinations of attribute levels, i.e. choice sets, to be tested in the experiment. In theory, all possible combinations of the attribute levels would be tested. However, doing so is impractical \cite{kuhfeld2005experimental}; it would be too expensive and place too much of a cognitive load on participants, likely resulting in poor data quality. To demonstrate, the number of possible combinations of the attribute levels for the uptake part was $108 (= 3^3 x 2^2)$ for the UK experiment and $192 (= 4^1 x 3^1 x 2^4)$ for Japan experiment. Hence, in line with general practice, a subset of all possible choice sets --- known as an \textit{orthogonal fractional factorial} design --- was used. 

For the Japan experiment which took place first, we considered three factors: statistical power, cognitive load, and budget constraint. In principle, the more choice sets and the higher the number of participants, the higher the statistical power. However, the higher the number of choice sets, the greater the cognitive load placed on participants; and the higher the number of participants, the more expensive the experiment. Soft-testing took place to test the cognitive workload of various numbers of choice sets with staff at the institution. The 36-choice set for each part of the experiment: the uptake and the actual uses of the app, blocking into 4 versions, was concluded as a suitable design. In the UK study, we considered the same factors and used the insights gained from the previous design from the Japan study. However, we also took into account the fact that participants would be members of the public; hence could be less patient with the 36-choice sets design. The 8-choice set, blocking into 2 versions, was chosen. The choice sets for both studies were selected randomly from all possible combinations of choice sets using SAS JMP.

\subsubsection{Data collection}

\begin{figure}[bp]
    \centering
    \begin{subfigure}\footnotesize[a]{ A choice set for the uptake of a VPN app}
        \centering
        \includegraphics[width=0.60\linewidth]{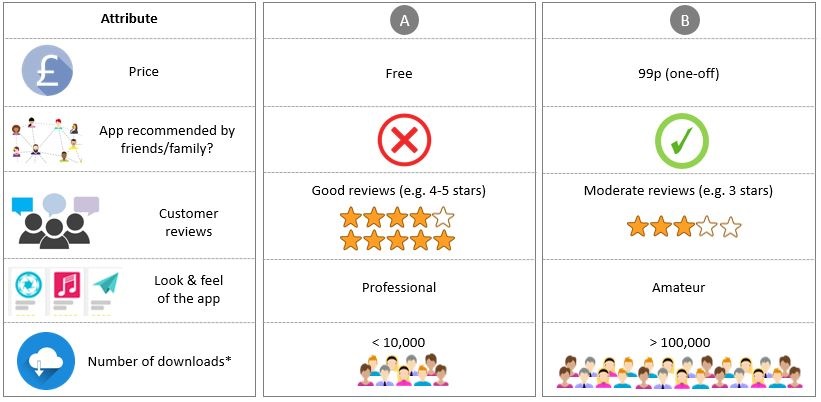}
    \end{subfigure}%
    ~
    \\
    \begin{subfigure}\footnotesize[b]{ A choice set for the actual uses of a VPN app}
        \centering
        \includegraphics[width=0.60\linewidth]{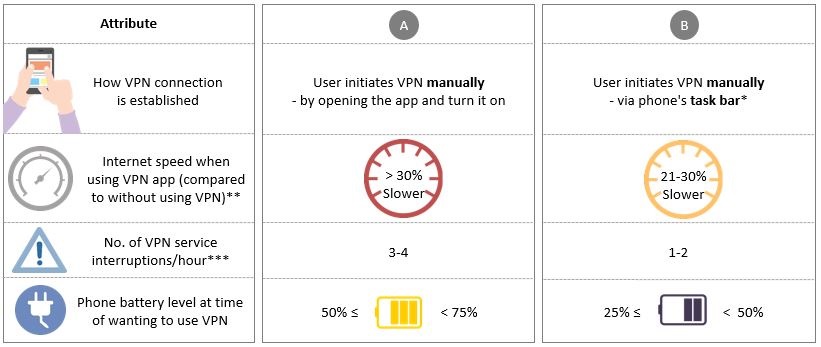}
    \end{subfigure}
    \caption{An example of a choice set for the UK experiment}
    \label{fig:Fig1}
\end{figure}

\begin{figure}[bp]
    \centering
    \begin{subfigure}\footnotesize[a]{ A choice set for the uptake of a VPN app}
        \centering
        \includegraphics[width=0.60\linewidth]{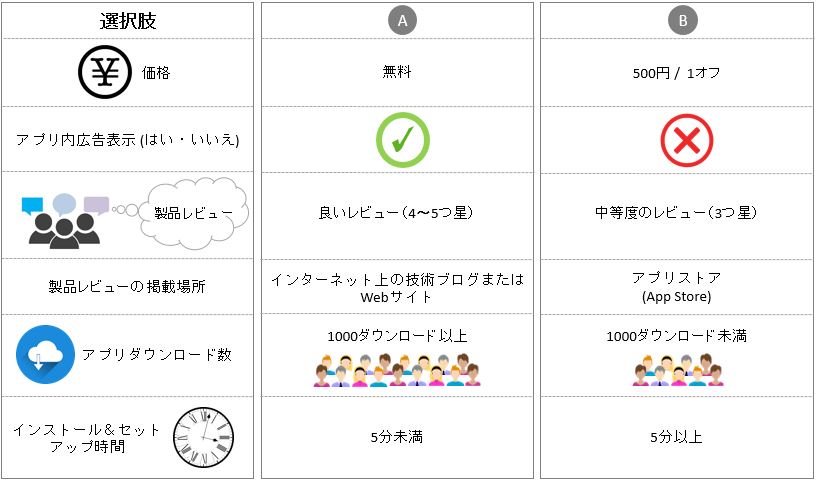}
    \end{subfigure}%
    ~
    \\
    \begin{subfigure}\footnotesize[b]{ A choice set for the actual uses of a VPN app}
        \centering
        \includegraphics[width=0.60\linewidth]{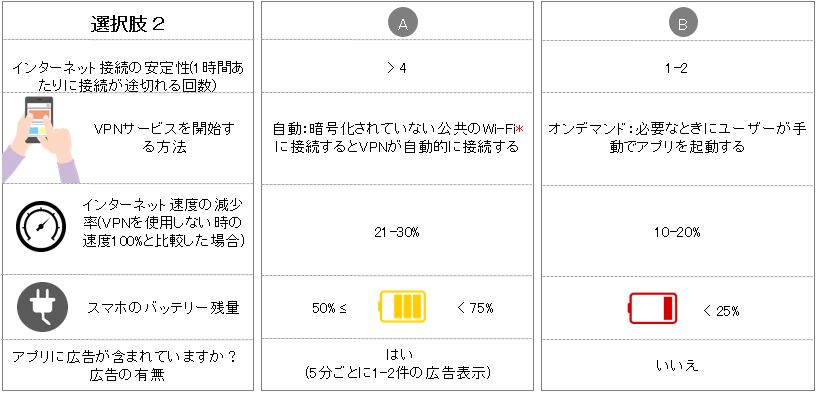}
    \end{subfigure}
    \caption{An example of a choice set for Japan experiment}
    \label{fig:Fig2}
\end{figure}

\paragraph{Experiment structure} We used LimeSurvey as a platform for our online experiment. Before starting the experiment, we provided participants (on-screen) with info about VPN and a short video clip of how it helped to mitigate the risks of using public Wi-Fi (in English and in Japanese). The experiment consisted of three parts. Part I set the scene by asking participants demographic questions, their usage and perceived risks of public Wi-Fi, and prior experience with VPN. Part II and III were the actual choice experiment for the uptake and the actual uses of a VPN app, respectively. In each part, participants were presented with a series of choice sets. Each choice set consisted of two competing two hypothetical VPN apps having the same set of attributes but with at least one (or more) attribute level(s) different from each other. The user interface design was localized for each country to make the experiment more engaging (Examples in Fig.~\ref{fig:Fig1} (UK) and Fig.~\ref{fig:Fig2} (JP)). Participants were asked to choose the app they preferred. No personal identifiable information (PII) was collected --- hence data were anonymous. We pilot tested the system before launching it.

\paragraph{Recruitment} In the UK, we recruited participants via Prolific Academic. In Japan, we were allowed to advertise our study via student and staff mailing lists and verbally in classrooms, and put up flyers advertising our studies at one of the participating institutes. In both countries, eligible participants were restricted to individuals living in the UK/Japan, at least 18 years old, and used public Wi-Fi at least from time to time. Each participant in the UK was awarded \pounds{3} (for a 15-min experiment). In Japan, each participant at two participating institutions was awarded a \textyen1000 gift voucher (for a 1-hour experiment). Participants at another institution, however, were recruited on a voluntary basis as monetary payments was not allowed.

\subsubsection {Data cleansing and analysis}
Participant responses were refined to optimize data quality. Incomplete records or records which failed the fatigue test were removed. After data cleansing, we had 243 responses (149 UK, 94 JP) for Part I of the experiment and 239 responses (148 UK, 91 JP) for Part II from mixed demographic (Appendix: Table ~\ref{table:table7}). Data were analyzed using SAS JMP Choice Model suite. We analyzed the \textit{main effects} of the choice outcome using the likelihood ratio (LR) tests. Next, we used the Effect Marginal function to analyze the marginal probabilities and marginal utilities for each \textit{main effects}. The Probability Profiler function was used to compare choice probabilities among potential combinations of attribute level\footnote{Defined as $\exp{(U)}/(\exp{(U)}+\exp{(U_{b})})$ where $U$ is the utility for the current settings and $U_{b}$ is the utility for the baseline settings; implies that the probability for the baseline settings is 0.5 \cite{sas}.} and to identify a set of attribute levels that would return maximized desirability i.e. the ideal VPN app that participants perceived as most desirable. Finally, the WTP function was used to estimate participants' Willingness to Pay (WTP) for a VPN app given a change in certain attribute values. 

\subsection {Ethics consideration}
We submitted the study design to the IRB of the institution in the UK. The application covered both the UK and Japan study. We were granted permission for the study provided that we: 1) informed participants about the study, 2) explained the study to the participants and received consent from them prior to data collection, 3) where PII was collected, complied with applicable data protection laws, and 4) delete any PII upon the publication of the study. We also obtained approval to run the study from the institutions in Japan.

\section{Results} \label{sec:results}

\subsection{Attributes affecting the VPN uptake}
\subsubsection{Main effects}
The app review rating --- a form of herding attitude describing the tendency for people to follow others \cite{baddeley2018copycats,baddeley2017behavioural} --- exerted the most influence (UK: ($x^2(2) = 564.98$, $p < 0.0001$; JP: ($x^2(2) = 717.17$, $p < 0.0001$)). Price followed (UK: ($x^2(2) = 169.60$, $p < 0.0001$); JP: ($x^2(3) = 607.94$, $p < 0.0001$)). In the UK, the decisions were also influenced by recommendation by friends ($x^2(1) = 87.85$, $p < 0.0001$) and the number of app downloads($x^2(2) = 22.63$, $p < 0.0001$) but to a lesser extent than the app review rating and the price. However, user interface of a VPN app did not significantly affect participants' decisions ($x^2(1) = 0.12$, $p > 0.05$). In Japan, the presence of an in-app ads ($x^2(1) = 15.57$, $p < 0.0001$), source of app review rating ($x^2(1) = 15.01$, $p < 0.001$), and the number of downloads of the app ($x^2(1) = 10.40$, $p < 0.001$) also played a significant role but to a lesser extent than the app review rating and the price. Table ~\ref{table:table2} shows a summary of the \textit{main effects} on the uptake decisions. 

\begin{table}[tp]
\caption{\textit{Main effects} on the uptake decisions}
{\scriptsize
\centering
\begin{tabular}[left]{|l|c|c|} 
 \hline
 \textbf{Attribute} & \textbf{UK} ($n=149$) & \textbf{JP} ($n=94$) \\
 \hline
 App review rating &564.98(2)*** &717.17(2)***\\
 \hline
 Price &169.60(2)*** &607.94(3)***\\
 \hline
 No. of downloads &22.63(2)*** &10.40(1)**\\
 \hline
 Interface (UI) &0.12(1) &n/a \\
 \hline
 Friend recommendation &87.85(1)*** &n/a\\
 \hline
 Source of app review &n/a &15.01(1)**\\
 \hline
 In-app ads display &n/a &15.57(1)***\\
 \hline
 Installation and setup time &n/a &0.07(1)\\ [1ex]
 \hline
 \end{tabular}
 \\ \tiny \begin{flushright} () degree of freedom, ***, **, * significant at $p < 0.0001, 0.001, 0.05$ \end{flushright}
 }
\label{table:table2}
\end{table}

\subsubsection{Effect marginal}
The marginal utility (MU) showed that participants preferred a free VPN app over a paid app ($MU = 1.15$ (UK), $= 0.86$ (JP)). All other attribute levels being equal, the marginal probability (MP) of participants choosing a free app was $0.65$ (UK) and $0.48$ (JP). Participants also preferred an app with a good review rating over a moderate and bad rating ($MU = 1.80$ (UK), $= 0.77$ (JP)). The MP of any participant choosing an app with a good review rating, all other attribute levels being equal, was $0.82$ (UK) and $0.58$ (JP). The MU and MP for all attributes are in Fig.~\ref{fig:Fig4} in the Appendix.

\subsubsection{Probability profiler}
Review rating and price were found most influential.

\paragraph{Review Rating} In the UK, all other attribute levels being equal, a change in the app rating from moderate to good increased the probability of uptake by $0.33 (= 0.83-0.50$ probability of choosing the UK baseline app\footnote{Set as a free app (\pounds{0}), moderate app review rating, not referred by friends, had $< 10,000$ download, and had an amateur-looking interface}). A downward change to a bad review, however, reduced the probability by $0.40 (= 0.10-0.50$). Similarly, in Japan, a change from moderate to good increased the probability of uptake by $0.14 (=0.64-0.50$ probability of choosing the baseline app\footnote{Set as a free app (\textyen0), moderate app review rating (based on info from App store), had $<1,000$ downloads, had in-app ads, and required $< 5$ mins installation}). A downward change to a bad review reduces the probability by $0.26 (= 0.50-0.24$). Again, the results underline the importance of the herding attitude in security decisions. Our participants followed the crowd too when deciding whether to download and install a VPN app, just like many ordinary decisions in life \cite{baddeley2018copycats}.

\paragraph{Price} An increase in price reduced the probability of uptake in a linear manner. In the UK, introducing a \pounds{0.99} fee to a baseline free app reduced the probability by $0.18 (= 0.32-0.50)$. A more expensive option of \pounds{4.99}/month drove the probability down by $0.44 (= 0.06-0.50)$. In Japan, likewise, the probability was reduced by $0.12 (= 0.38-0.50)$, $0.27 (= 0.23-0.50)$, and $0.35 (= 0.15-0.50)$ if charging \textyen100/one off, \textyen500/one off, and \textyen1000/month.

\subsubsection{Willingness-to-Pay (WTP)} Even though price was a deterrent, participants were willing to make a trade-off and pay for a VPN app if some attribute levels were to change. All other attributes levels of a baseline app being equal, participants in the UK and Japan were willing to pay \pounds{3.05} ($SE = 0.44$) and \textyen343 ($SE = 34.00$) if the baseline free app with moderate rating had a good review rating. The UK participants were also willing to pay \pounds{2.05} ($SE = 0.35$) if the baseline app was recommended by friends, and pay \pounds{1.19} ($SE = 0.46$) if the baseline free app (having less than 10K downloads) had a number of downloads between 10K and 100K. Similarly, participants in Japan were willing to pay \textyen91 ($SE = 26.60$) if the baseline free app (having less than 1K downloads) had more than 1K downloads. However, they were not willing to pay for an app in order to remove in-app ads ($WTP = -106.19$, $SE = 25.98$), suggesting that participants in Japan would rather download and install a free VPN app with ads. 

\subsubsection{Maximized desirability} The maximized desirability calculation showed some similarities in the ideal sets of attribute levels that participants in both countries viewed as most desirable. In the UK, the ideal attribute set ($Desirability = 0.80$ (on a scale of 0 to 1), $Utility = 3.80$ ($min = 3.12$, $max = 4.48$)) was observed in a free VPN app with a good review, recommended by friends, 10K -- 100K downloads, and with an amateur look and feel. In Japan, the ideal set ($Desirability = 0.80$, $Utility = 1.88$ ($min = 1.72$, $max = 2.04$)) was also a free VPN app with a good review rating (based on the info on app store), but also with an installation time of less than 5 mins, a greater than 1K download, and with the presence of in-app ads.   

\subsection{Attributes affecting the actual uses of a VPN app}
\subsubsection{Main effects}
Participants in the UK and in Japan prioritized the attributes affecting the actual uses of a VPN app differently. In the UK, Internet speed when using VPN played the most significant role ($x^2(2) = 262.96$, $p < 0.0001$), followed by battery level on mobile devices at the time of wishing to use VPN ($x^2(3) = 126.05$, $p < 0.0001$). Connection stability (i.e. the number of dropped connections/hour) also significantly influenced the decisions but to a lesser extent ($x^2(2) = 32.44$, $p < 0.0001$). However, in Japan, connection stability played the most significant role ($x^2(2) = 140.82$, $p < 0.0001$), followed by battery level on mobile devices ($x^2(3) = 132.56$, $p < 0.0001$). The decision to use a VPN was also affected by whether the app displayed an ad ($x^2(2) = 81.14$, $p < 0.0001$). Internet speed when using VPN, however, did not affect the decision to use a VPN app as much as it did to the UK participants ($x^2(2) = 22.47$, $p < 0.0001$).

In both countries, the method to initiate a VPN app --- whether automatic or manual --- did not significantly affect the decision to use the app (UK: $x^2(2) = 1.29$, $p > 0.05$; JP: ($x^2(1) = 1.52$, $p > 0.05$)). This suggested that the endowment effect --- the tendency for people to generally value something more once they own it \cite{kahneman1991anomalies} --- may not apply to a VPN app. Table ~\ref{table:table4} provides a summary of the \textit{main effects} on the decisions to actually use a VPN app.

\begin{table}[tp]
\caption{\textit{Main effects} on the decisions to use a VPN app}
{\centering
\scriptsize
\begin{tabular}[p]{|p{5cm}|c|c|} 
 \hline
 \textbf{Attribute} & \textbf{UK} ($n=148$) & \textbf{JP} ($n=91$) \\
 \hline
 Battery level &126.05(3)*** &132.56(3)***\\
 \hline
 Internet speed when using VPN &262.96(2)*** &22.47(2)***\\
 \hline
 Connection stability &32.44(2)*** &140.82(2)***\\
 \hline
 Method to initiate VPN &1.29(2) &1.52(1) \\
 \hline
 In-app ads display & n/a &81.14 (1)***\\
 \hline
 \end{tabular}
 \\ \tiny \begin{flushright} () degree of freedom, ***, **, * significant at $p < 0.0001, 0.001, 0.05$ \end{flushright}
 }
\label{table:table4}
\end{table}

\subsubsection{Effect marginal} 
In the UK, all other attribute levels being equal, participants preferred a 10-20\% decrease in Internet speed when using VPN ($MU = 0.94$, $MP = 0.65$), rather than the two other slower attribute levels. For mobile phone battery level at the time of wishing to use VPN, the 75-100\% level was the most preferred choice ($MU = 0.70$, $MP = 0.43$), all other attribute levels being equal. The battery level of less than 25\% was the least preferred option ($MU = -0.93$, $MP = 0.08$), suggesting influence from resource preservation heuristic. In Japan, all other attribute levels being equal, participants preferred 1-2 dropped connections/hour when using VPN, the lowest among the three attribute levels ($MU = 0.35$, $MP = 0.46$). Participants also preferred no in-app ads displayed when using a VPN app ($MU = 0.20$, $MP = 0.60$). Similar to the UK, the 75-100\% battery level was the most preferred choice ($MU = 0.35$, $MP = 0.34$) whilst the 25\% battery level was the least preferred choice ($MU = -0.46$, $MP = 0.15$), suggesting that resource preservation heuristic was universal in nature. The MU and MP for all attributes are in Fig.~\ref{fig:Fig7} in the Appendix.

\subsubsection{Probability profiler} Statistically significant results from the Internet speed, connection stability, and battery levels were observed.

\paragraph{Internet speed} In the UK, where participants were most concerned about the Internet speed, the probability of using the app reduced by $0.12 (= 0.28-0.50)$ and $0.36 (= 0.14-0.50)$ if the speed was reduced from the baseline of 10-20\% slower to 21-30\% slower and to $>$30\% slower, respectively. This suggests that stabilising Internet speed when VPN is in use is needed. 

\paragraph{Connection stability} In Japan, where participants were most concerned about VPN connection stability, the probability of using the app reduced by $0.07 (= 0.43-0.50)$ if the number of dropped connections/hr changed from 3-4 times/hr to $>$4 times/hr. However, if it changed to only 1-2 times/hr, the probability of using the app increased by $0.09 (= 0.59-0.50)$, suggesting the need to minimise interruptions to the service to encourage users to use a VPN app.

\paragraph{Battery level on mobile device} All other attribute levels being equal, the probability of participants using a VPN app decreased as the battery level decreased in both countries, but to a lesser extent in Japan. In the UK, the probability reduced by $0.10$ ($= 0.40-0.50$ probability of choosing the baseline app\footnote{Set as an app with 3-4 dropped connections/hr, 10-20\% slower in Internet speed (than without VPN), users needed to initiate the app manually (via app), and there was 75-100\% battery level left on a user's mobile device}), $0.18 (= 0.32-0.50)$, and $0.34 (= 0.16-0.50)$ if the battery level was to reduce from 75-100\% to 50-74\%, 25-49\%, and $<$ 25\%, respectively. In Japan, when the battery level was depleted from 75\%-100\% to 50-75\%, and 25-50\%, the probability of using a VPN app decreased by $0.07 (= 0.43-0.50$ probability of choosing the baseline app\footnote{Set as an app with an in-app ads displayed, 3-4 dropped connections/hr, 10-20\% slower in Internet speed (than without VPN), users needed to initiate VPN manually, and there was 75-100\% battery level on a user's mobile}). If the battery was less than 25\%, the probability of using the app decreased by $0.19 (= 0.31-0.50)$. One possible explanation for this difference is that carrying power banks when out and about is more common in Japan. This finding supports evidence from previous studies showing how the resource preservation heuristic affects risk-mitigating decisions; users were also reluctant to update software due to fear of draining their mobile phone battery \cite{vaniea2016tales,ndibwile2018smart4gap}.

\subsubsection{Maximized desirability}
The ideal sets of attribute levels that participants in both countries viewed as most desirable were fairly similar. In the UK, the ideal set ($Desirability=1.00$ (on a scale of 0 to 1), $Utility=1.99$ ($min=1.70$, $max=2.27$)) was observed in a VPN app having 1-2 dropped connections/hr, being 10-20\% slower in Internet speed (compared to without VPN), connecting automatically when using public Wi-Fi, and with participants having 75-100\% battery level on mobile devices at the time of wishing to use VPN. The same set of levels, plus having no in-app ads, was found to be most desirable in Japan ($Desirability=0.91$, $Utility=1.03$ ($min=0.90$, $max=1.17$)). 

\section{Discussion} \label{sec:discuss}
Our study provides three key insights. First, several attributes of a VPN app significantly affected the decisions to download and install and to actually use the app. Second, preferences for some of these attributes were driven by biases in decision-making, specifically the herding attitude and the resource preservation heuristic. Third, the preferences for and the priority given to these attributes were not always universal. These insights offer a number of potential applications for VPN providers, public policy makers, and cyber security research community.

\subsection{VPN app providers}
First, although price significantly affected the app uptake decisions, contrary to conventional wisdom, it was not the most important factor. Rather, the review rating of the app was. Participants were willing to pay for a free VPN app if the review rating was 4-5 stars. This finding suggests that VPN app providers should address customer feedback promptly to increase/maintain the review rating. Next, the findings that the installation and setup time, and the look and feel of the app did not significantly affect the uptake decisions should be welcoming to VPN app providers. From an economics standpoint, app developers can spend less time on perfecting these attributes, reducing the overall costs of development. Moreover, for Japan in particular, the findings that participants were not willing to pay to remove in-app ads would help to guide VPN providers to plan pricing more carefully. Hence, the pay-to-remove-ads strategy, as seen in many apps today, is unlikely to be attractive for VPN users in Japan. VPN providers can also use the insights to develop a VPN app that is attractive to use and drive users to use it as a habit. These include several proposals. First, a VPN app should consume minimal battery power because the battery preservation heuristic significantly deter the desire to use the app. Minimising the number of dropped connections and stabilising Internet speed when VPN is being used are other attributes that VPN providers should consider improving. 

\subsection{Public policy makers}
Public policy makers can use the insights from the study to develop attractive awareness campaigns to promote VPN adoption. An awareness campaign which utilizes the power of social influence to change behavior could be more effective than just giving out general messages about VPN e.g., showing how many people have already downloaded VPN apps could potentially attract interest from the public. Studies in behavioral economics (e.g., \cite{asch1956studies,nolan2008normative}) have shown that this \enquote*{social nudging} technique works, albeit with different products/services.

\subsection{Cyber security research community}
The finding that preferences for some of the attributes of a VPN app were driven by biases in decision-making --- the herding attitude and the resource preservation heuristic --- is beneficial for the study of security decisions. It emphasizes that security decisions too were affected by biases. However, empirical evidence to support this notion is still limited. Further exploring of these biases is needed to help us understand security decisions better. The community would also benefit from adapting the DCEs used in this study to other security contexts. 

\section{Limitations and future work} \label{sec:limit}
Our study has limitations. First, there could be other attributes of a VPN affecting the decisions to adopt a VPN app but were not tested in our DCEs. However, we believe that our approach to attribute identification was sufficiently rigorous and that our guided questions and the interview probing techniques adequately addressed these issues. Preferences for VPN attributes could also be driven by the \textit{subject effects} --- factors pertaining to individuals e.g., gender, age, perceived risks of public Wi-Fi; we seek to explore them in details in future work. Next, in the choice experiment, despite providing clear instructions and adequate information about VPN, and using engaging experiment design, some participants may not have paid full attention. However, our pilot tests and fatigue tests were designed to detect these potential pitfalls. 

There were also uncommon threats to the external validity of the results. Our evidence were from the UK and Japan; both are developed economies with good Internet infrastructure. Users or potential users of a VPN app in other countries may have different preferences e.g., price may be the most critical factor in developing economies. Next, in Japan, participants were recruited from participating institutions only. Their knowledge of and experience with VPN and cyber security in general were likely to be higher than that of the general public. Future studies can also adapt and improve upon our method to cyber security study using DCEs. The DCEs can also be applied to other cyber security contexts such as investigating factors affecting the adoption of other security tools.

\section{Conclusion} \label{sec:conclusion}
We investigated attributes affecting user decision to adopt a VPN app, a tool which helps to mitigate the privacy and security risks when using unsecured networks such as public Wi-Fi. The novelty of this study lies in it being the first to examine the attributes of a VPN using DCEs and drawing cross-cultural evidence from the UK and Japan. Our findings showed that various attributes of a VPN app can be designed to drive the uptake and the actual usage of the app. The latter, in particular, is a difficult challenge. Asking people to form a new habit is hard but we showed that --- with the right incentives --- it is not entirely hopeless. We also showed that preferences for and priorities given to certain VPN app's attributes are not universal, suggesting that a customized VPN app for different markets would be more favourable than the one-size-fit-all app, mostly seen in the app store today. Moreover, we provided another evidence that security decisions --- in the VPN adoption context --- were affected by biases commonly observed in decisions-making in general too.  

\section{Acknowledgement} \label{sec:ack}
This work was funded by the ICS-CoE Core Human Resources Development Program and the EU TEAM Erasmus Mundus scholarship. We thank Caroline Wardle, Kuniko Kumano, Takahashi Hideaki, Shane Johnson, and iPLab members for their help, and Lyn Lua and Yvette Vermeer for reviewing the paper.

\bibliographystyle{splncs04}
\bibliography{reference}

\clearpage
\appendix
\section{Appendix}
\begin{table}[h!]
\caption{Demographic of participants in the interviews}
\centering
\tiny
\begin{tabular}[p]{|p{5cm}|p{.5cm}|p{.5cm}|p{.5cm}|p{.5cm}|}

 \hline
 \multirow{2}{6em}{\textbf{Demographic}} &\multicolumn{4}{|c|}{\textbf{Country}}\\
 \cline{2-5}
  &\multicolumn{2}{|c|}{\textbf{\textit{UK}}} &\multicolumn{2}{|c|}{\textbf{\textit{JP}}}\\
 \cline{1-5}
 \hline
 \textbf{\textit{Gender}} &\textbf{\textit{n}} &\% &\textbf{\textit{n}} &\%\\
 \hline
 Female &8 &53 &3 &18\\
 \hline
 Male &7 &47 &14 &82\\
 \hline
 Total  &15 &100 &17 &100\\
 \hline
 \textbf{\textit{Education}} &\textbf{\textit{n}} &\% &\textbf{\textit{n}} &\%\\
 \hline
 A Level or vocational training &4 &27 &1 &6\\
 \hline
 Bachelor's degree &5 &33 &10 &59\\
 \hline
 Postgraduate's degree &6 &40 &6 &35\\
 \hline
 Total  &15 &100 &17 &100\\
 \hline
 \textbf{\textit{Age}} &\textbf{\textit{n}} &\% &\textbf{\textit{n}} &\%\\
 \hline
 18-25 &8 &53 &9 &53\\
 \hline
 26-35 &3 &20 &4 &24\\
 \hline
 36-45 &3 &20 &2 &12\\
 \hline
 56-65 &1 &7 &2 &12\\
 \hline
 Total  &15 &100 &17 &100\\
 \hline

\end{tabular}
\label{table:table6}
\end{table}

\begin{table}[h!]
\centering
\caption{Demographic of participants in the DCE}
\tiny
\begin{tabular}[p]{|p{5cm}|p{.5cm}|p{.5cm}|p{.5cm}|p{.5cm}|}
 \hline
 \multirow{2}{6em}{\textbf{Demographic}} &\multicolumn{4}{|c|}{\textbf{Country}}\\
 \cline{2-5}
  &\multicolumn{2}{|c|}{\textbf{\textit{UK}}} &\multicolumn{2}{|c|}{\textbf{\textit{JP}}}\\
 \cline{1-5}
 \hline
 \textbf{\textit{Gender}} &\textbf{\textit{n}} &\% &\textbf{\textit{n}} &\%\\
 \hline
 Female &80 &54 &9 &10\\
 \hline
 Male &67 &45 &83 &88\\
 \hline
 Prefer not to say &2 &1 &2 &2\\
 \hline
 Total  &149 &100 &94 &100\\
 \hline
 \textbf{\textit{Education}} &\textbf{\textit{n}} &\% &\textbf{\textit{n}} &\%\\
 \hline
 GCSE Level (or equivalent) &13 &9 &nil &nil\\
 \hline
 A Level (or equivalent) &34 &23 &5 &5\\
 \hline
 Diploma/vocational training &22 &15 &1 &1\\
 \hline
 Bachelor's degree &60 &40 &44 &47\\
 \hline
 Postgraduate's degree &20 &13 &44 &47\\
 \hline
 Total  &149 &100 &94 &100\\
 \hline
 \textbf{\textit{Age}} &\textbf{\textit{n}} &\% &\textbf{\textit{n}} &\%\\
 \hline
 18-25 &41 &28 &65 &69\\
 \hline
 26-35 &50 &34 &18 &19\\
 \hline
 36-45 &37 &25 &8 &9\\
 \hline
 46-55 &14 &9 &3 &3\\
 \hline
 56-65 &6 &4 &nil &nil\\
 \hline
 66+ &1 &1 &nil &nil\\
 \hline
 Total  &149 &100 &94 &100\\
 \hline
 \textbf{\textit{Employment}} &\textbf{\textit{n}} &\% &\textbf{\textit{n}} &\%\\
 \hline
 Not working - Fulltime students &19 &13 &49 &52\\
 \hline
 Not working - others &16 &11 &2 &2\\
 \hline
 Not working - permanently sick/disable &6 &4 &nil &nil\\
 \hline
 Working - full time &82 &55 &25 &27\\
 \hline
 Working - part time &19 &13 &18 &19\\
 \hline
 Total  &149 &100 &94 &100\\
 \hline
 \textbf{\textit{Income}} &\textbf{\textit{n}} &\% &\textbf{\textit{n}} &\%\\
 \hline
 Up to \pounds{12,500} &87 &58 &nil &nil\\
 \hline
 \pounds{12,501} to \pounds{50,000} &12 &8 &nil &nil\\
 \hline
 \pounds{50,001} to \pounds{150,000} &50 &34 &nil &nil\\
 \hline
 Under \textyen1,950,000 &nil &nil &66 &70\\
 \hline
 \textyen1,950,000 to \textyen3,300,000 &nil &nil &5 &5\\
 \hline
 \textyen3,300,000 to \textyen6,950,000 &nil &nil &16 &17\\
 \hline
 \textyen6,950,000 to \textyen9,000,000 &nil &nil &4 &4\\
 \hline
 \textyen9,000,000 to \textyen18,000,000 &nil &nil &3 &3\\
 \hline
 Total  &149 &100 &94 &100\\
 \hline
\end{tabular}
\label{table:table7}
\end{table}

\begin{figure}
\hfill
\subfigure[Main effect marginal -- UK]{\includegraphics[width=5.5cm]{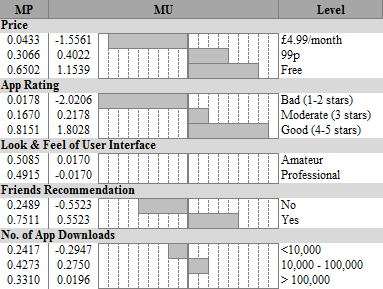}}
\hfill
\subfigure[Main effect marginal -- JP]{\includegraphics[width=5.5cm]{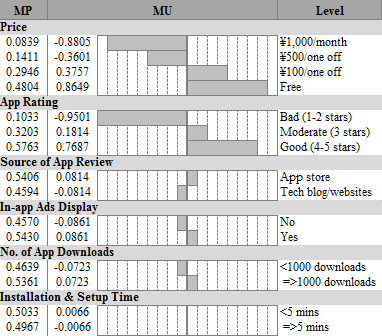}}
\hfill
\caption{The main effects' marginal probability and utility for the uptake decisions}
\label{fig:Fig4}
\end{figure}

\begin{figure}
\hfill
\subfigure[Main effect marginal -- UK]{\includegraphics[width=5.5cm]{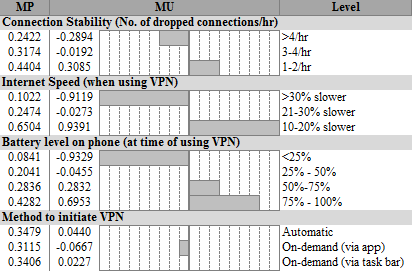}}
\hfill
\subfigure[Main effect marginal -- JP]{\includegraphics[width=5.5cm]{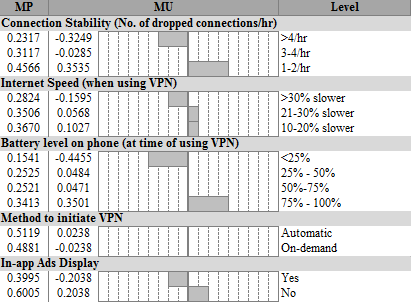}}
\hfill
\caption{The main effects' marginal probability and utility for the decisions to use a VPN app}
\label{fig:Fig7}
\end{figure}

\end{document}